\documentclass[aps,rmp,twocolumn]{revtex4}
\usepackage{graphicx}
\usepackage{amsmath}
\makeatletter
\@ifxundefined\bibnotes@sw{\@booleantrue\bibnotes@sw}{}%
\bibpunct{}{}{,}{s}{}{\textsuperscript{,}}%
\def\@onlinecite#1{\begingroup\let\@cite\NAT@citenum\citealp{#1}\endgroup}%
\appdef\class@documenthook{%
 \@ifxundefined\place@bibnumber{%
  \let\place@bibnumber\place@bibnumber@sup
 }{}%
}%
\makeatother
\begin{document}
\title{Fundamental Aspects of Quantum Brownian Motion}
\author{Peter H{\"a}nggi}
\author{Gert-Ludwig Ingold}
\affiliation{Institut f{\"u}r Physik, Universit{\"a}t Augsburg,
             86135 Augsburg, Germany}
%\date{\today}
\begin{abstract}
With this work we elaborate on the physics of quantum noise in thermal
equilibrium and in stationary non-equilibrium. Starting out from the celebrated
quantum fluctuation-dissipation theorem we discuss some important consequences
that must hold  for open, dissipative quantum systems in thermal equilibrium.
The issue of quantum dissipation is exemplified with the fundamental problem of
a damped harmonic quantum oscillator. The role of quantum fluctuations is
discussed in the context of both, the nonlinear generalized quantum Langevin 
equation and the path integral approach. We discuss the consequences of the 
time-reversal symmetry for an open dissipative quantum dynamics and, 
furthermore,  point to a series of subtleties and possible pitfalls. The path 
integral methodology is applied to the decay of metastable states assisted by 
quantum Brownian noise.
\end{abstract}

\maketitle
\begin{bfseries}
This work deals with the description of quantum Brownian motion in linear and
nonlinear quantum systems that exhibit frictional influences. The symmetries of
thermal equilibrium impose severe constraints on the time evolution properties
of open quantum systems. These lead to a quantum generalization of the
classical Einstein relation that connects friction with the strength of thermal
quantum fluctuations. There exist a variety of theoretical roadways to model
quantum dissipation. Here, we discuss the topic for the prominent case of a
damped harmonic oscillator upon combining thermodynamics and linear
response theory. A dissipative nonlinear quantum dynamics can be dealt with a
generalized quantum Langevin equation, a path integral formulation, or in terms 
of a generalized quantum master equation for the corresponding reduced 
dynamics. We illustrate the situation for the problem of the dissipative decay 
out of a metastable state. Furthermore, we point out a series of subtleties, 
pitfalls and shortcomings that one must be aware of when confronted with the 
world of quantum  noise driven phenomena.
\end{bfseries}

\section{Introduction}
Albert Einstein  explained the phenomenon of Brownian motion in his
\textit{annus mirabilis} of 1905 by use of statistical methods which he
ingeniously combined with the laws of thermodynamics \cite{einstein1905}.
In this pioneering work he as well provided a first link between the dissipative
forces and the impeding thermal fluctuations, known as the \textit{Einstein
relation} which relates the strength of diffusion to the friction. This
intimate connection between dissipation and related fluctuations was put
on a firm basis much later when Nyquist \cite{nyquist1928} and Johnson
\cite{johnson1928} considered the spectral density of voltage- and
current-fluctuations.

What role do quantum mechanics and the associated quantum fluctuations
play in this context? After the birth of quantum mechanics in the early 1920's 
we can encounter in the very final paragraph of the 1928 paper by Nyquist
for the first time the introduction of quantum mechanical noise via the 
substitution of the energy $kT$ from the classical equipartition law
\cite{note1} by the thermally averaged quantum energy (but leaving out the 
zero point energy contribution) of the harmonic oscillator. Nyquist's remark 
thus constitutes a precursor of the celebrated work by Callen and Welton 
\cite{callenwelton1951} who generalized the relations by Einstein, Nyquist and 
Johnson to include quantum effects: In their work they put forward a generally 
valid connection between the response function and the associated quantum 
fluctuations in equilibrium, the quantum fluctuation-dissipation theorem.

Without  doubt, quantum fluctuations constitute a prominent noise source in
many nano-scale and biological systems. For example, the tunnelling and the
transfer of electrons, quasi-particles, and alike, is assisted by noise for
which the quantum nature \textit{cannot} be neglected. The features of this
noise change drastically as a function of temperature: At sufficiently high
temperatures a crossover does occur to classical Johnson-Nyquist noise.

With this work we shall present various methods and schemes of modelling
quantum Brownian motion from first principles. In particular, the thermal noise 
must at all times obey the quantum version of the fluctuation-dissipation
theorem (\`a la Callen-Welton). This latter property is necessary in order
to be consistent with the second law of thermodynamics and the principle of
(quantum) detailed balance. We elaborate on several alternative but equivalent
methods to describe quantum noise and quantum Brownian motion per se:
These are the functional integral method for dissipative quantum systems
\cite{grabertetal88, dittrich88} and  time-dependent driven quantum systems
\cite{grifoni98}, the quantum Langevin (operator) approach \cite{QLE},
stochastic schemes \cite{nelson66,GHT79}, or the concept of stochastic
Schr\"odinger equations \cite{SS}. In doing so, we call attention to distinct
differences to the classical situation and, as well, identify a series of
delicate pitfalls which must be observed when making even naive looking
approximations. Such pitfalls involve, among others, the rotating wave
approximation, the use of quasi-classical Langevin forces, the quantum
regression hypothesis and/or the Markov approximation
\cite{grabertetal88, talkner86, grifoni98}.

\section{The quantum fluctuation-dissipation theorem and its implications}

As already mentioned, in 1951 Callen and Welton proved a pivotal
relation between equilibrium fluctuations and dissipative transport
coefficients. Note also that this quantum fluctuation-dissipation relation
holds true independent of particle statistics. The following cornerstone
achievements can be found in this primary work \cite{callenwelton1951}:
\begin{itemize}
\item The generalization of the classical Nyquist's formula to the quantum
      case.
\item The quantum mechanical proof that susceptibilities are related to the
      spectral densities of symmetrized correlation functions.
\end{itemize}

For a single degree of freedom, \textit{linear} response theory yields for the
change of the expectation value of an operator-valued observable $B$ due to
the action of a (classical) force $F(t)$ that couples to the conjugate
dynamical operator $A$
\begin{equation}
\label{response}
{<}\delta B(t){>} = \int_{-\infty}^{t}ds \chi_{BA}(t-s) F(s)\,.
\end{equation}
Here, $\delta B(t)=B(t)-{<}B{>}_0$ denotes the difference with respect to
the thermal equilibrium average ${<}B{>}_0$ in the absence of the force.
The reaction of the system is contained in the response function $\chi_{BA}(t)$
with a so-called dissipative part
\begin{equation}
\chi_{BA}^d(t)= \frac{1}{2i}[\chi_{BA}(t) - \chi_{AB}(-t)]\,.
\end{equation}
The Fourier transform of $\chi_{BA}^d(t)$ will be denoted by 
$\tilde\chi_{BA}^d(\omega)$. It is worth noting here that only when $A=B$ does 
this part in fact coincide with the imaginary part of the complex-valued 
susceptibility $\tilde\chi_{BA}''(\omega)$.

The fluctuations are described by the equilibrium correlation function
\begin{equation}
C_{BA}(t) = {<}\delta B(t)\delta A(0){>}_\beta
\end{equation}
at inverse temperature $\beta = 1/kT$. The correlation function is
complex-valued because the operators $B(t)$ and $A(0)$ in general do not
commute. While the antisymmetric part of $C_{BA}(t)$ is directly related 
to the response function by linear response theory, the power spectrum of
the \textit{symmetrized} correlation function
\begin{equation}
\label{eq:sba}
S_{BA}(t)=\frac{1}{2}{<}\delta B(t)\delta A(0)+\delta A(0)\delta B(t){>}
\end{equation}
depends on the Fourier transform of the dissipative part of the response
function via
\begin{equation}
\label{eq:fdt}
S_{BA}(\omega) = \hbar\coth\left(\frac{\hbar\omega}{2kT}\right)
\tilde\chi_{BA}^d(\omega)\,.
\end{equation}
This result is the quantum version of the fluctuation-dissipation theorem
as it relates the fluctuations described by $S_{BA}(\omega)$ to the
dissipative part $\tilde\chi_{BA}^d(\omega)$ of the response.

In the spirit of the work by Nyquist and Johnson we consider as an example
the response of a current $\delta I$ through an electric circuit subject to a
voltage change $\delta V$. This implies $B=I$ and, because the voltage couples
to the charge $Q$, $A=Q$. The response of the circuit is determined by
$\delta I(\omega)=Y(\omega) \delta V(\omega)$ where the admittance $Y(\omega)$
is identical to the susceptibility $\chi_{IQ}(\omega)$. As a consequence of
$I=\dot Q$, the symmetrized power spectrum of the current fluctuations is
given by $S_{II}(\omega)=i\omega S_{IQ}(\omega)$ so that we obtain
\begin{align}
\label{eq:FDTI}
S_{II} (\omega) &= \hbar\omega\coth\left(\frac{\hbar\omega}{2kT}\right)
\mathrm{Re}Y(\omega)\\
&=2\left[\frac{\hbar\omega}{2}+\frac{\hbar\omega}{e^{\beta\hbar\omega}-1}
\right]\text{Re}Y(\omega)\,.\nonumber
\end{align}

In the high temperature limit $kT \gg \hbar\omega$, we recover the results of
Nyquist and Johnson, i.e.\ $ S_{II}(\omega)\rightarrow 2kT \mathrm{Re}
Y(\omega)$. For the Markovian limit of an ohmic resistor, where
$Y(\omega)=1/R$ this result simplifies to read $S_{II}(\omega)= 2kT/R$.
The quantum version was already anticipated by Nyquist in the last
paragraph of his 1928 paper \cite{nyquist1928}. However, he made use of the
original expression of Planck which yields only the second contribution
present in the lower line of (\ref{eq:FDTI}). Nyquist thus missed the first
term arising from the vacuum energy which already appears in a paper by
Planck published in 1911.\cite{planck11}

On the other hand, in the extreme quantum limit $kT\ll\hbar\omega$, we find
that $S_{II}(\omega) \rightarrow \hbar\omega \mathrm{Re}Y(\omega)$.
In particular, this implies that at \textit{zero frequency} the spectral
weight of the current fluctuations vanishes in the generic case where
the admittance does not exhibit an infrared divergence.

We cannot emphasize enough that the quantum fluctuation-dissipation relation 
(\ref{eq:fdt}) and corresponding implications hold true for any isolated, closed
quantum system. Thus, upon contracting the dynamics in full phase space onto a
reduced description of  an open quantum system exhibiting dissipation these
relations hold true nevertheless. Therefore, care must be taken when invoking
approximations in order to avoid any violation of these rigorous relations. We
next consider the role of quantum dissipation for an exactly solvable
situation: the damped quantum harmonic oscillator dynamics.

\section{Quantum dissipation: the damped harmonic oscillator}
\label{sec:dampedosc}

\subsection{Equilibrium correlation functions}
Let us next consider the most fundamental case of a simple open quantum
system, namely the damped harmonic oscillator. This problem could be
tackled by setting up a microscopic model describing the coupling to
environmental degrees of freedom to which energy can be transferred
irreversibly, thus giving rise to dissipation. Such an approach will be
introduced in Sect.~\ref{subsec:bath}. On the other hand, the linearity of
the damped harmonic oscillator allows us as alternative to proceed on a
phenomenological level. This approach is closely related to the usual classical
procedure where damping is frequently introduced by adding in the equation of
motion a force proportional to the velocity.

Classically, the motion of a harmonic oscillator subject to linear friction
is determined by
\begin{equation}
M\ddot q +M\int_{-\infty}^t ds\gamma(t-s)\dot q(s)+M\omega_0^2 q=0\,.
\label{eq:eqomdho}
\end{equation}
In the example of an electric circuit mentioned in the previous section, a
damping kernel $\gamma(t)$ with memory would correspond to a
frequency-dependent admittance. In the special case of ohmic friction
corresponding to $Y(\omega)=1/R$, the damping force is proportional to
the velocity of the harmonic oscillator, so that the equation of motion
reads
\begin{equation}
M\ddot q+M\gamma\dot q+M\omega_0^2q=0\,.
\label{eq:eqomohmic}
\end{equation}
In (\ref{eq:eqomdho}) and (\ref{eq:eqomohmic}) the mass, frequency, and 
position of the oscillator are denoted by $M$, $\omega_0$, and $q$, 
respectively. Due to the Ehrenfest theorem, the equation of motion 
(\ref{eq:eqomdho}) is still valid in the quantum regime if we replace $q$ by 
its expectation value. As a consequence, the quantum mechanical dynamic 
susceptibility agrees with the classical expression \cite{GWT84,RHW85}
\begin{equation}
\chi_{qq}(\omega)=\frac{1}{M}\frac{1}{-\omega^2-i\omega\tilde\gamma(\omega)
+\omega_0^2}
\label{eq:chiqq}
\end{equation}
where $\tilde\gamma(\omega)$ denotes the Fourier transform of the damping
kernel $\gamma(t)$.

As mentioned before, the response function directly yields the
antisymmetric part of the position autocorrelation function $C_{qq}(t)$. It
therefore suffices to discuss the symmetrized part $S_{qq}(t)$ defined
according to (\ref{eq:sba}). Furthermore, our linear system with linear
damping represents a stationary Gaussian process so that all higher order
correlation functions may be expressed in terms of second order correlation
functions.\cite{GWT84} In addition, equilibrium correlation functions 
containing momentum operators $p$ can be reduced to position correlation 
functions by means of $p=M\dot q$. The dynamics of the damped harmonic 
oscillator can therefore entirely be described by the response function, 
i.e.\ the Fourier transform of (\ref{eq:chiqq}) and
$S_{qq}(t)$.\cite{GWT84,RHW85,grabertetal88}

In the case of ohmic damping, $\tilde\gamma(\omega)=\gamma$, the position
autocorrelation can be explicitly evaluated from the
fluctuation-dissipation theorem (\ref{eq:fdt}). The inverse Fourier
transform into the time domain is determined by the poles on the right-hand
side. The dissipative part of the dynamic susceptibility leads to four poles
at $\omega= \pm(\bar\omega\pm i\gamma/2)$ with
$\bar\omega=(\omega_0^2- \gamma^2/4)^{1/2}$ which contribute to the correlation
function $S_{qq}(t)$ at all temperatures. At sufficiently low temperatures,
the poles of the hyperbolic cotangent at $\omega=\pm i\nu_n$ with the
Matsubara frequencies $\nu_n=2\pi n/\hbar\beta$ become important as well.
After performing the contour integration in (\ref{eq:fdt}), one arrives 
at\cite{GWT84,RHW85,grabertetal88}
\begin{align}
S_{qq}(t)&=\frac{\hbar}{2M\bar\omega}\exp\left(\displaystyle-\frac{\gamma}{2}
\vert t\vert\right)\nonumber\\
&\qquad\times
\frac{\sinh(\hbar\beta\bar\omega)\cos(\bar\omega t)+\sin(\hbar\beta\gamma/2)
\sin(\bar\omega \vert t\vert)}{\cosh(\hbar\beta\bar\omega)-
\cos(\hbar\beta\gamma/2)}\nonumber\\
&\quad-\frac{2\gamma}{M\beta}\sum_{n=1}^{\infty}\frac{\nu_n
\exp(-\nu_n \vert t\vert)}{(\nu_n^2+\omega_0^2)^2-\gamma^2\nu_n^2}.
\label{eq:sqq}
\end{align}

In the limit of high temperatures the second term vanishes and the first term
yields the classical correlation function. Quantum corrections to this term are
relevant at temperatures of the order $\hbar\omega_0/k$ or below, and these
corrections may be obtained from weak coupling theories like the quantum master
equation approach \cite{grabert82,haake73,spohn80,alicki87}. However, there is
another regime at temperatures below $\hbar\gamma/4\pi k$. Here, the second
term may initially be small, but nevertheless it may dominate the long-time
behavior of the correlation function. This becomes particularly apparent in the
limit of zero temperature where the exponential functions in the second term in
eq.~(\ref{eq:sqq}) sum up to an algebraic long-time behavior, i.e.\
$S_{qq}(t)=-(\hbar\gamma/\pi M\omega_0^4)t^{-2}$. Its relevance for the
dynamical evolution of the damped harmonic oscillator depends on the details of
the initial preparation \cite{lnm90}. Although the algebraic decay results from
the zero temperature limit it can also be observed at low, but finite
temperatures during intermediate times before an exponential decay with time
constant $\nu_1$ sets in \cite{jung85}. The occurrence of additional time
scales besides $\gamma$ at low temperatures leads to shortcomings with the
quantum regression hypothesis and allows for the decay of correlations on time
scales longer than $\gamma$.\cite{talkner86}

\subsection{The reduced density matrix and the partition function}
In the previous section, we have seen that the dynamics of a damped
harmonic oscillator can be fully described in terms of the position
autocorrelation function (\ref{eq:sqq}) and its time derivatives as well as
the (classical) response function. If one is interested only in equilibrium
expectation values of arbitrary operators acting in the Hilbert space of the
harmonic oscillator, it is sufficient to know the reduced density matrix.
By means of arguments analogous to the dynamic case presented in the previous 
section, the reduced density matrix can only depend on second moments of 
position and momentum, ${<}q^2{>}_\beta$ and ${<}p^2{>}_\beta$, respectively. 
The equilibrium density matrix then necessarily takes the form
\cite{grabertetal88}
\begin{align}
\label{eq:reddm}
\rho_\beta(q,q') &= \frac{1}{(2\pi{<}q^2{>})^{1/2}}\\
&\qquad\times\exp\left[-\frac{(q+q')^2}{8{<}q^2{>}_\beta}
   -\frac{{<}p^2{>}_\beta}{\hbar^2}(q-q')^2\right]\,.\nonumber
\end{align}

The second moments are found to read
\begin{equation}
{<}q^2{>}_\beta = \frac{1}{M\beta}\sum_{n=-\infty}^{+\infty}
\frac{1}{\omega_0^2+\nu_n^2+\vert\nu_n\vert\hat\gamma(\vert\nu_n\vert)}
\end{equation}
and
\begin{equation}
{<}p^2{>}_\beta = \frac{M}{\beta}\sum_{n=-\infty}^{+\infty}
\frac{\omega_0^2+\vert\nu_n\vert\hat\gamma(\vert\nu_n\vert)}
{\omega_0^2+\nu_n^2+\vert\nu_n\vert\hat\gamma(\vert\nu_n\vert)}\,,
\label{eq:p2}
\end{equation}
where we have introduced the Laplace transform of the damping kernel
\begin{equation}
\hat\gamma(z) = \int_{0}^{\infty} dt \exp(-zt) \gamma(t)\,.
\end{equation}

We note that for strictly ohmic damping the second moment of the momentum
(\ref{eq:p2}) exhibits a logarithmic divergence which can be removed by
introducing a finite memory to the damping mechanism. For finite coupling to
the environment, i.e.\ for finite damping strength $\gamma$, the reduced
density matrix (\ref{eq:reddm}) obviously does not agree with the canonical
density matrix $\exp(-\beta H_S)$ at the same temperature, where $H_S$
denotes the Hamiltonian of the undamped harmonic oscillator.

\begin{figure}
\includegraphics[width=0.8\columnwidth]{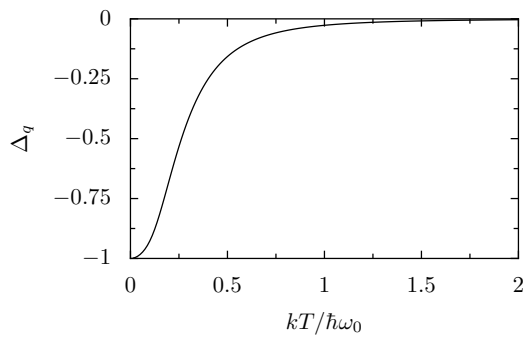}
\caption{The weak coupling correction $\Delta_q$ to ${<}q^2{>}_\beta$
according to (\protect\ref{eq:deltaq}) is depicted as a function of the
temperature $T$. For $kT\gg\hbar\omega_0$, the correction becomes
negligible.}
\label{fig:corr}
\end{figure}

In order to get an idea of the deviation of the true reduced density matrix
from the canonical one, we consider the leading corrections to the second
moment of the position due to the finite coupling to the environment.
Expanding in orders of the damping strength $\gamma$, we obtain for ohmic
damping
\begin{equation}
\frac{{<}q^2{>}_\beta(\gamma)}{{<}q^2{>}_\beta(\gamma=0)} =
1+\frac{\gamma}{\pi\omega_0}\Delta_q+O(\gamma^2)
\end{equation}
with
\begin{equation}
\label{eq:deltaq}
\Delta_q = \frac{\hbar\beta\omega_0}{2\pi}
\frac{\text{Im}\psi'\left(\displaystyle i\frac{\hbar\beta\omega_0}{2\pi}\right)}
{\coth\left(\displaystyle\frac{\hbar\beta\omega_0}{2}\right)}\,.
\end{equation}
Here, $\psi'$ denotes the first derivative of the digamma function. The
correction $\Delta_q$ is depicted in Fig.~\ref{fig:corr} as a function of
temperature. We find that the leading corrections are particularly
important in the quantum regime, $kT\ll\hbar\omega_0$, while in the
classical regime the corrections to the canonical density matrix are
negligible.

As we have already mentioned, a finite memory time of the damping kernel
or, equivalently, a finite cutoff frequency $\omega_D$ for the environmental
mode spectrum is needed in order to keep the second moment of the momentum
(\ref{eq:p2}) finite. If $\omega_D\gg\omega_0,\gamma$, the corrections to
the canonical density matrix for weak coupling will only be small if the
temperature is larger than the cutoff frequency, i.e. $kT\gg\hbar\omega_D$.

The differences between the correct reduced density matrix (\ref{eq:reddm})
and the canonical one are also reflected in the partition function. Without
specifying a microscopic model for the environment, the partition function
$\mathcal{Z}$ for the damped harmonic oscillator can be obtained by the
requirement that it generates the second moment of position according to
\cite{GWT84}
\begin{equation}
{<}q^2{>}_\beta =
-\frac{1}{M\beta\omega_0}\frac{d}{d\omega_0}\ln(\mathcal{Z})\,.
\end{equation}
This leads to the product representation of the partition function, i.e.
\begin{equation}
\mathcal{Z} = \frac{1}{\hbar\beta\omega_0}\prod_{n=1}^{\infty}
\frac{\nu_n^2}{\nu_n^2+\nu_n\hat\gamma(\nu_n)+\omega_0^2}\,.
\label{eq:partfunc}
\end{equation}

\begin{figure}
\includegraphics[width=0.8\columnwidth]{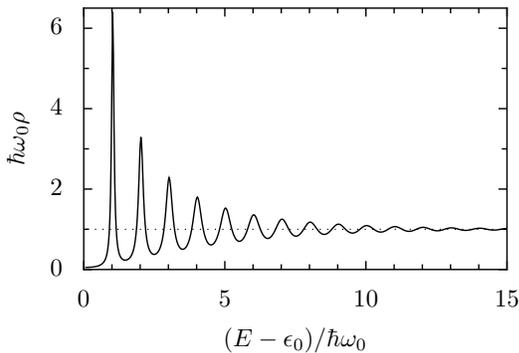}
\caption{The density of states defined by inversion of the relation
(\protect\ref{eq:zbrhob}) for a weakly damped harmonic oscillator with
$\gamma=0.1\omega_0$ exhibits broadened peaks close to the energies
$\epsilon_0+n\hbar\omega_0$. A delta function at the ground state energy
$\epsilon_0$ is not shown explicitly. The dashed line represents the
average density of states.}
\label{fig:dos}
\end{figure}

The properties of this partition function become more transparent if one
relates it to a density of states $\rho(E)$ according to \cite{hanke95}
\begin{equation}
\label{eq:zbrhob}
\mathcal{Z}(\beta) = \int_0^{\infty}dE\rho(E)\exp(-\beta E)\,.
\end{equation}
The factor $(\hbar\beta\omega_0)^{-1}$ in (\ref{eq:partfunc}) can then be
interpreted in terms of the average density of states
$(\hbar\omega_0)^{-1}$ indicated in Fig.~\ref{fig:dos} as dotted line. We
further note that the partition function diverges for purely ohmic damping.
However, it can be shown that this divergence is entirely due to a
divergence of the ground state energy $\epsilon_0$ in the presence of ohmic
dissipation \cite{ingold02}. For large cutoff frequencies, the poles of the
partition function, which determine the density of states, can then be
determined from the condition $\nu_n^2+\gamma\nu_n+\omega_0^2=0$. These
poles give rise to a density of states which for weak damping exhibits
narrow peaks whose width is in agreement with the result from Fermi's
golden rule. Fig.~\ref{fig:dos} depicts an example for $\gamma/\omega_0=0.1$.
In view of the remark made before, the density of states is shifted by the
ground state energy. In addition, a delta peak at the ground state energy
has been omitted. With increasing damping strength, the peaks broaden so that
for sufficiently strong damping a rather featureless density of states
results which decreases with increasing energy to the average density of
states (cf.\ Fig.~3 in Ref.~\onlinecite{hanke95}).

\section{Dissipation in nonlinear quantum systems: the generalized quantum 
Langevin equation (QLE)}
\label{sec:qmdissipative}

\subsection{Bath of oscillators}
\label{subsec:bath}
For nonlinear systems the arguments given in the previous section no longer
apply. In particular, second-order correlation functions are not sufficient
anymore to completely describe the damped system. An alternative approach to
quantum dissipative systems starting from a Hamiltonian at first sight does
not seem feasible because the absence of time-dependent forces implies energy 
conservation. However, as we will see below, once it is realized that 
dissipation arises from the coupling to other degrees of freedom, it is 
straightforward to model a damped quantum system in terms of a Hamiltonian.

A well known technique to describe a statistical dynamics governed by
fluctuations is given by the method of generalized master equations and the
methodology of generalized Langevin equations. This strategy is by now well
developed for \textit{thermal equilibrium systems}. Here the projector operator
methodology \cite{mori65,grabert82,haake73, spohn80, alicki87} yields a
clear-cut method to obtain the formal equations, either for the rate of change
of the probability, i.e.\ the \textit{generalized quantum master equation}
(QME) or the (generally nonlinear) \textit{generalized quantum Langevin
equation} (QLE).

Already for the case of relaxation towards a unique thermal equilibrium
specified by a single temperature $T$, the equivalence between the two
approaches is not very transparent \cite{grabertetal80}. A crucial role is
played by the fluctuational force which explicitly enters the equivalence,
such as corresponding cumulant averages to an arbitrary high order. This fact
is not appreciated generally, because one often restricts the discussion to the
first two cumulants only, namely the average and its auto-correlation. It is a
fact that little is known about the connection of the generalized master
equation and the corresponding generalized Langevin equation in a nonlinear 
situation \cite{grabertetal80,hanggi97}.

A popular model for the dynamics of a dissipative quantum system subject to 
quantum Brownian noise is obtained by coupling the system of interest to a bath 
of harmonic oscillators. Accordingly, we write for the total Hamiltonian
\begin{align}
\label{eq:hamiltonian}
H&=\frac{p^2}{2M}+V(q,t)\\
&\qquad+\sum_{i=1}^N\left[\frac{p_i^2}{2m_i}+
\frac{m_i}{2}\omega_i^2x_i^2-qc_ix_i+q^2\frac{c_i^2}{2m_i\omega_i^2}
\right]\nonumber
\end{align}
where the first two terms describe the system as a particle of mass $M$ moving
in a generally time-dependent potential $V(q,t)$. The sum contains the
Hamiltonian for a set of $N$ harmonic oscillators which are
\textit{bi-linearly} coupled with strength $c_i$ to the system. Finally, the
last term, which depends only on the system coordinate, represents a potential
renormalization term which is needed to ensure that $V(q,t)$ remains the bare
potential. This Hamiltonian has been studied since the early 60's for systems
which are weakly coupled to the environmental degrees of freedom
\cite{QLE,senit60,uller66,zwanz73,haake73,spohn80, alicki87, HTB90}. Only after 
1980, it was realized by Caldeira and Leggett \cite{caldeiraleggett83} that 
this model is also applicable to strongly damped systems and may be employed 
to describe, for example, dissipative tunnelling in solid state physics and 
chemical physics \cite{HTB90}.

One may convince oneself that the Hamiltonian (\ref{eq:hamiltonian}) indeed 
models dissipation. Making use of the solution of the Heisenberg equations of
motion for the external degrees of freedom \cite{note2} one derives a reduced
system operator equation of motion, the so-called \textit{generalized quantum 
Langevin equation} \cite{QLE}
\begin{equation}
M\ddot q (t)+M\int_{t_0}^{t} d s\gamma(t-s)\dot q(s)+\frac{dV(q,t)}{dq}
= \xi(t)
\label{eq:QLE}
\end{equation}
with the damping kernel
\begin{equation}
\gamma(t)= \gamma (-t)=\frac{1}{M}\sum_{i=1}^N\frac{c_i^2}{m_i\omega_i^2}\cos(\omega_i t)
\label{eq:dampingkernel}
\end{equation}
and the quantum Brownian force operator
\begin{align}
\label{eq:xi}
\xi(t)&=-M\gamma(t-t_0)q(t_0)\nonumber\\
&\quad+\sum_{i=1}^Nc_i\bigg(x_i(t_0)\cos(\omega_i [t-t_0])\\
&\qquad\qquad\qquad+\frac{p_i(t_0)}{m_i\omega_i}\sin(\omega_i[t-t_0])\bigg)\,.
\nonumber
\end{align}
The generalized quantum Langevin equation (\ref{eq:QLE}) appears first in a 
paper by Magalinski\u{\i} \cite{QLE} who started from (\ref{eq:hamiltonian}) 
in the absence of the potential renormalization term.

The force operator (\ref{eq:xi}) depends explicitly on the initial conditions 
at time $t_0$ of the bath position operators $x_i(t_0)$ and bath momenta 
$p_i(t_0)$. The initial preparation of the total system, which fixes the 
statistical properties of the bath operators and the system degrees of 
freeedom, turns the force $\xi(t)$ into a random operator. Note that this 
operator depends not only on the bath properties but as well on the initial 
system position $q(t_0)$. To fully specify the reduced dynamics it is thus of  
importance to specify the preparation procedure. This in turn then also fixes 
the statistical properties of the quantum Brownian noise. Clearly, in order to 
qualify as a stochastic force the random force $\xi(t)$ should not be biased; 
i.e.\ its average should be zero at all times. Moreover, this Brownian quantum 
noise should constitute a \textit{stationary} process  with time homogeneous
correlations. 

Let us also introduce next the auxiliary random force $\eta(t)$, defined by
\begin{equation}
\label{bathforce}
\eta(t) = \xi(t) + M\gamma(t-t_0)q(t_0)
\end{equation}
which only involves bath operators. In terms of this new random force the QLE 
(\ref{eq:QLE}) no longer assumes the form of an ordinary generalized Langevin 
equation: it now contains an inhomogeneous term $\gamma(t-t_0) q(t_0)$, the 
initial slip term \cite{hanggi97,ingold02}. This term is often neglected in the 
so-called ``Markovian limit'' when the friction kernel assumes the ohmic form
$\gamma(t)\rightarrow 2\gamma\delta(t)$. For a correlation-free preparation, 
the initial total density matrix is given by the product $\rho_T = 
\rho_S(t_0)\rho_\text{bath}$, where $\rho_S(t_0)$ is the initial system
density matrix.  The density matrix of the bath alone assumes canonical 
equilibrium, i.e.
\begin{equation}
\label{rhobath}
\rho_\text{bath} = \frac{1}{\mathcal{N}} \exp\left(-\beta\sum_{i=1}^N
\left[\frac{p_i^2}{2m_i}+\frac{m_i}{2}\omega_i^2x_i^2\right]\right)\,,
\end{equation}
with $\mathcal{N}$ denoting a normalization constant. 

The statistical properties of the random force $\eta(t)$ then follow 
immediately: $\eta(t)$ is a stationary \textit{Gaussian operator noise} obeying
\begin{align}
\label{eq:noiseaverage}
{<}\eta(t){>}_{\rho_\text{bath}}&= 0 \\
\label{eq:noisecorrelation}
S_{\eta\eta}(t-s)&=
\frac{1}{2}{<}\eta(t)\eta(s)+\eta(s)\eta(t){>}_{\rho_\text{bath}}\\
&=\frac{\hbar}{2}\sum_{i=1}^N\frac{c_i^2}{m_i\omega_i}
\cos\big(\omega_i(t-s)\big)\coth\left(\frac{\hbar\omega_i}{2kT}\right)\; .
\nonumber
\end{align}
Being an operator-valued noise, its commutator does not vanish
\begin{equation}
\label{noisecommutator}
[\eta(t),\eta(s)] = -i\hbar \sum_{i=1}^N\frac{c_i^2}{m_i\omega_i}
\sin\big(\omega_i(t-s)\big)\,.
\end{equation}

Setting for the initial position operator $q(t_{0})=q_0$, the last expression 
in (\ref{eq:noisecorrelation}) is also valid for the noise correlation 
$S_{\xi\xi}(t)$ of the noise force $\xi(t)$ provided the average is now
taken with respect to a bath density matrix which contains shifted oscillators.
The initial preparation of the bath is then given by the new density matrix
$\hat\rho_\text{bath}$;
\begin{align}
\label{eq:1.g}
\hat\rho_\text{bath} &= \frac{1}{\mathcal{N}} \exp \bigg\{-\beta \bigg[
\sum_{i} \frac{p^{2}_{i}}{2m_{i}}\\
&\qquad\qquad\qquad\qquad+\frac{m_{i}\omega^{2}_{i}}{2}
\left(x_{i} - \frac{c_{i}}{m_{i} \omega^{2}_{i}} \, q_0
\right)^{2} \bigg]\bigg\}\,.\nonumber
\end{align}

In some physical situations a microscopic model for the external degrees of
freedom is available \cite{HTB90,weiss93}. Examples are the electromagnetic
modes in a resonator acting as a reservoir or the dissipation arising from
quasi-particle tunnelling through Josephson junctions\cite{esa84}. In the case
of an electrical circuit containing a resistor one may use the classical
equation of motion to obtain the damping kernel and model the environment
accordingly. This approach has been used e.g.\ to model Ohmic dissipation in
Josephson junctions in order to study its influence on tunnelling processes
\cite{schoen90}, and to describe the influence of an external impedance in
the charge dynamics of ultrasmall tunnel junctions \cite{ingol92}.

This scheme of the QLE can also be extended to the nonequilibrium case with the 
system attached to two baths of different temperature \cite{zurchertalkner90}. 
Two most recent applications address the problem of the thermal conductance 
through molecular wires that are coupled to leads of different temperature. 
Then the heat current assumes a form similar to the Landauer formula for
electronic transport: The heat current is given in terms of a transmission 
factor times the difference of corresponding Bose functions \cite{segaletal03}.

Furthermore, the  QLE concept can also be extended to {\it fermionic} systems 
coupled to electron reservoirs and which, in addition, may be exposed to 
time-dependent driving \cite{camaletetal}. The corresponding Gaussian quantum 
noise is now composed of fermion annihilation operators.

\subsection{Consequences of time-reversal symmetry}
Let us now discuss some further properties of this QLE. If the potential 
$V(q,t)$ in (\ref{eq:hamiltonian}) does \textit{not} explicitly depend on time 
$t$, the dynamics of the full Hamiltonian (\ref{eq:hamiltonian}) obeys time 
reversal symmetry. It is thus an immediate consequence that the reduced 
dynamics must be invariant under time reversal as well. This must hold true 
despite the fact that the QLE has been constructed to allow for a description 
of quantum dissipation. It is thus instructive to see how the validity under 
time reversal emerges from the contracted description in terms of
the QLE in (\ref{eq:QLE}).

Given the time of preparation $t_{0}$, reversing the time amounts to 
substituting time $t$ by $t_{0} - (t-t_{0})= 2t_{0} -t$. Using again the
random force $\eta(t)$ we can recast the QLE dynamics after the time reversal 
into the form
\begin{align}
\label{TRQLE}
&M\ddot q(2t_0-t)+M\int_{t_0}^{2t_0-t} ds\,\gamma(2t_0-t-s)\dot q(s)+
\frac{dV(q)}{dq}\nonumber\\
&\qquad\qquad= \xi(2t_0-t) \nonumber \\
&\qquad\qquad= \eta(2t_0-t) -\gamma(2t_0-t-t_0)q(t_0)\,.
\end{align}

Setting next $x(t)=q(2t_0-t)$ and observing that $\dot x(t)= -\dot q(2t_0-t),
\ddot x(t)=\ddot q(2t_0-t)$, we find after the substitution of the integration 
time $u = 2t_0-s$ from (\ref{TRQLE}) the result
\begin{align}
\label{TRQLE2}
&M\ddot x(t)+M\int_{t_0}^{t} d u\,\gamma(u-t)\dot x(u)+\frac{dV(q)}{dq}
\nonumber\\
&\qquad\qquad= \xi(2t_0-t) \nonumber  \\
&\qquad\qquad= \eta(2t_0-t) -\gamma(t_{0}-t)x(t_0)\,.
\end{align}

Noting that the damping kernel is an even function of its argument,
$\gamma(u-t)=\gamma(t-u)$, and that
$x(t_0)=q(t_0)$, we find upon changing all signs of the initial momenta
$p_i(t_0)\rightarrow -p_i(t_0)$ for the noise forces the relations 
$\eta(2t_0-t)= \eta(t)$ and $\xi(2t_0-t)=\xi(t)$. We conclude that the time 
reversed motion $x(t)=q(2t_0-t)$ indeed obeys again a QLE of the form 
(\ref{eq:QLE}). This even holds true in the Markovian limit where $\gamma(t-s)= 
2\gamma \delta(t-s)$ as one can convince oneself by smearing out the delta
function symmetrically. The QLE then reads for all times $t$
\begin{equation}
M\ddot q(t) + \text{sgn}(t-t_0) M\gamma \dot q(t) + \frac{dV(q)}{dq}
= \xi(t)\,,
\end{equation}
where $\text{sgn}(x)$ denotes the sign of $x$.

The dissipation is reflected by the fact that for times $t>t_0$ the reduced
dynamics for $q(t)$ exhibits a damped (quantum)-behavior on a time scale given
by the Poincar\'e recurrence time \cite{hemmeretal58,mazur60}; the latter 
reaches essentially infinity for all practical purposes if only the bath
consists of a sizable number of bath oscillator degrees of freedom 
\cite{HTB90,hemmeretal58}.

\subsection{Subtleties and pitfalls}
The use of the generally nonlinear QLE (\ref{eq:QLE}) is limited in 
practice for several reasons. Moreover, the  application of the QLE bears some
subtleties and pitfalls which must be observed when making approximations.
Some important features are:
\begin{itemize}
\item The QLE (\ref{eq:QLE}) is an operator equation that acts in the full
      Hilbert space of system and bath. The coupling between system and
      environment also implies an entanglement upon time evolution even for the
      case of an initially factorizing full density matrix. Together with the
      commutator property of quantum Brownian motion, see 
      eq.~(\ref{noisecommutator}), we find that the reduced, dissipative 
      dynamics of the position operator $q(t)$ and momentum operator $p(t)$ 
      obey the Heisenberg uncertainty relation for all times.
\end{itemize}

This latter feature is crucial. For example, the non-Markovian (colored)
Gaussian quantum noise with real-valued correlation $S_{\xi\xi}(t)=
S_{\xi\xi}(-t)$ cannot simply be substituted by a \textit{classical}
non-Markovian Gaussian noise force which identically obeys the correlation
properties of (Gaussian) quantum noise $\xi(t)$. An approximation of this type
clearly would not satisfy the commutator property for position and conjugate
momentum of the system degrees of freedom.

The literature is full of various such attempts wherein one approximates the
quantum features by corresponding colored classical noise sources, e.g.\ see
Refs.~\onlinecite{adelman76,schmid82,ray}. Such schemes work at best near a
quasi-classical limit \cite{schmid82, eckern90}, but even then care must be
exercised. For example, for problems that exhibit an exponential sensitivity,
such as the dissipative decay of a metastable state discussed in the next 
section, such an approach gives no exact agreement with the quantum dissipative 
theory \cite{HTB90,weiss93}. It is only in the classical high temperature 
limit, where the commutator structure of quantum mechanics no longer
influences the result. Perfect agreement is only achieved in the classical 
limit.

The study of quantum friction in a nonlinear quantum system by means of the QLE 
(\ref{eq:QLE}) is plagued by the fact that the nonlinearity forbids an explicit 
solution. This solution, however, is needed to obtain the statistical 
properties such as mean values and correlation functions. This (unknown) 
nonlinear response function also determines the derivation of the rate of 
change of the reduced density operator, i.e.\ the QME, and its solution of the open quantum system.

The very fact that the QLE acts in \textit{full} Hilbert space of system and
environment also needs to be distinguished from the classical case of a
generalized Langevin equation. There, the stochastic dynamics acts solely on 
the state space of the system dynamics with the (classical) noise properties 
specified a priori \cite{hanggithomas82}.
\begin{itemize}
\item The quantum noise correlations can, despite the explicit microscopic
      expression given in (\ref{eq:noisecorrelation}), be expressed solely 
      by the macroscopic friction kernel $\gamma(t)$.
\end{itemize}
This result follows upon noting that the Laplace transform $\hat\gamma(z)$ of
the macroscopic friction assumes with ${\rm Re} z >0$ the form
\begin{equation}
\label{Laplacegamma}
\hat\gamma(z) =\frac{1}{2 M}\sum_{i=1}^N\frac{c_i^2}{m_i\omega_i^2}
\left[\frac{1}{z-i\omega_i} + \frac{1}{z+i\omega_i}\right]\,.
\end{equation}
With help of the well known relation $1/(x+i0^+) = P(1/x)-i\pi\delta(x)$ we
find that
\begin{multline}
\label{Regamma}
\text{Re}\hat\gamma(z=-i\omega+0^+)\\
=\frac{\pi}{2 M}\sum_{i=1}^N\frac{c_i^2}{m_i\omega_i^2}
\left[\delta(\omega-\omega_i)+\delta(\omega+\omega_i)\right]\,.
\end{multline}
By means of (\ref{eq:noisecorrelation}) we then find the useful relation
\begin{align}
\label{eq:QEinsteinFDT}
&S_{\xi\xi}(t) = S_{\eta\eta}(t)\\
&\quad=\frac{M}{\pi}\int_{0}^{\infty} d\omega
\text{Re} \hat\gamma(-i\omega+0^+)
\hbar\omega \coth\left(\frac{\hbar\omega}{2kT}\right)\cos(\omega t)\,.
\nonumber
\end{align}

In the classical limit this relation reduces, independent of the preparation of
the bath with $\rho$ or $\hat\rho$, to the non-Markovian Einstein relation
$S_{\xi\xi}(t)= MkT\gamma(t)$. The relation (\ref{eq:QEinsteinFDT}) is by no 
means obvious: It implies that a modelling of quantum dissipation is possible 
in terms of macroscopic quantities such as the friction kernel $\gamma(t)$ and 
the temperature $T$. For other coupling schemes between system and bath we 
generally can no longer express the correlation of quantum noise exclusively in 
terms of macroscopic transport coefficients. As an example we mention the 
coupling of the system to a bath of two-level systems (spin bath) rather than 
to a bath of harmonic oscillators \cite{shaohanggi98}.

Note also the following differences to the classical situation of a
generalized Langevin equation.
\begin{itemize}
\item The quantum noise $\xi(t)$ is correlated with the initial position
      operator $q(t_0)$. \cite{schramm85} This feature that
      ${<}q(t_0)\xi(t){>}_{\hat\rho} \neq 0$ follows from the explicit form
      of the quantum noise $\xi(t)$. The correlation function vanishes
      only in the classical limit. Note also that the expectation value of the
      system-bath interaction is finite at zero temperature. These features 
      reflect the fact that at absolute zero temperature the coupling
      induces a non-vanishing decoherence via the zero-point fluctuations.
\end{itemize}
Moreover,
\begin{itemize}
\item the initial slip term $\gamma(t-t_0)q(t_0)$ appears also in the absence
      of the potential renormalization in the Hamiltonian 
      (\ref{eq:hamiltonian}). With this initial value contribution being 
      absorbed into the quantum fluctuation $\xi(t)$, these become stationary  
      fluctuations with respect to the initial density operator of the bath 
      $\hat\rho_{\text bath}$ given by (\ref{eq:1.g}). Note, however, that  
      with respect to an average over the bare, non-shifted bath density
      operator $\rho_{\text bath}$, the quantum fluctuations $\xi(t)$ would 
      become non-stationary.
\end{itemize}
It is also worthwhile to point out here that this initial value term in the 
QLE  should not be confused with the initial value term that enters the 
corresponding QME \cite{grabert82, haake73}.
In the case of a classical reduced dynamics it is always possible
-- by use of a corresponding projection operator -- to formally eliminate
this initial, inhomogeneous contribution in the generalized master equation
\cite{hanggithomas82,GTH77}. This in turn renders the time evolution of the 
reduced probability a truly linear dynamics. This property no longer holds for 
the reduced quantum dynamics \cite{nonlinearQD04}: For a non-factorizing 
initial preparation of system and bath this initial value contribution
in the QME generally is finite and  presents a true nonlinearity for the
time evolution law of the open quantum dynamics!

There exist even further subtleties which are worthwhile to point out. The
friction enters formally the QLE just in the same way as in the classical
generalized Langevin equation.
In particular, a time-dependent potential $V(q,t)$ leaves this friction kernel
invariant in the QLE. In contrast to the classical Markovian case, however,
where the friction enters the corresponding Fokker-Planck dynamics independent
of the time scale of driving, this is  no longer valid for the generalized
quantum master equation dynamics of the corresponding reduced density matrix.
\cite{zerbehanggi95,kohler97}

The solution of the QLE involves the explicit time-dependence of both the
friction and the potential forces. These in turn determine the statistical
properties of the density matrix. As a consequence, the friction force enters
the QME in a rather complex manner. This can already be verified
explicitly for a parametric dissipative oscillator dynamics, where the
time-dependent driving enters the diffusive kinetic evolution law of the
reduced density operator or its equivalent Wigner transform
\cite{zerbehanggi95,kohler97}.

For the bilinear system-bath interaction with the bath composed of harmonic
oscillators it was possible to integrate out the degrees of freedom of the bath
explicitly. Does this hold as well for other interactions? The
elimination of the bath degrees of freedom is still possible for a nonlinear
coupling to a bath of harmonic oscillators if the system part of the
coupling is replaced by a nonlinear operator-valued function of either the 
momentum or position degree of freedom of the system as long as the bath 
degrees of freedom appear linearly. The resulting friction
kernel then appears as a nonlinear friction but the influence of the bath
degrees of freedom still is obtained in exact form \cite{hanggi97}.

Yet another situation for which one can derive an exact QLE is when
a nonlinear system, such as a spin degree of freedom, interacts with a
collection of quantum (Bose) oscillators in such a way that the 
interaction Hamiltonian \textit{commutes} with the system Hamiltonian, thus
constituing a quantum non-demolition interaction. This case corresponds to 
pure dephasing and was addressed by {\L}uczka for the problem of a
spin in contact with a thermal heat bath \cite{luczka90}. It has since been
rederived many times, see e.g.\ Ref.~\onlinecite{vankampen95}.

We end this subsection by mentioning also the coupling of a system to a bath
of independent fermions with infinitely many excitation energies. A suitable
transformation then allows to map the dissipation onto a bosonic environment
with an appropriate coupling strength \cite{HTB90,camaletetal,fermionbath}.

\section{Path integrals and effective action}
\label{sec:pathintegrals}

\subsection{Nonlocal effective action}
A most effective approach to describe dissipation is based on the path integral
formulation of quantum mechanics \cite{feynmanrmp48}. In the path integral
formulation of quantum mechanics the propagator is expressed as
\begin{equation}
\langle q_f\vert\exp\left(-\frac{i}{\hbar}Ht\right)\vert q_i\rangle=
\int_{\displaystyle q(0)=q_i}^{\displaystyle q(t)=q_f}{\cal D}q
\exp\left(\frac{i}{\hbar} S[q]\right)
\label{eq:rtpi}
\end{equation}
where the integral runs over all possible paths starting at $q_i$ and
ending after time $t$ at $q_f$. The paths are weighted with a phase
factor which contains the classical action $S[q]$.

For the description of quantum dissipative systems it is important to realize
the analogy between the propagator and the equilibrium density matrix. The
latter is obtained by replacing $t$ by $-i\hbar\beta$. We thus obtain from
(\ref{eq:rtpi}) the path integral representation of the equilibrium density
matrix
\begin{equation}
\rho_{\beta}(q,q')=\frac{1}{Z_{\beta}}\int_{\displaystyle q(0)=q'}
^{\displaystyle q(\hbar\beta)=q}{\cal D}q\exp\left(-\frac{1}{\hbar}
S^E[q]\right)\,,
\label{eq:itpi}
\end{equation}
where $Z_{\beta}$ is the partition function. This integral is called
imaginary-time path integral in contrast to the real-time path integral
(\ref{eq:rtpi}). Note that in (\ref{eq:itpi}) the action $S[q]$
has been replaced by the so-called Euclidean action $S^E[q]$ which
is obtained by changing the sign of the potential term as a consequence
of the transition to imaginary times. In imaginary time we therefore have to
consider the motion in the inverted potential.

The connection between classical and quantum mechanics becomes particularly
apparent in the path integral formulation. The dominant contribution to the
integrals in (\ref{eq:rtpi}) and (\ref{eq:itpi}) arise from the stationary
points of the action, i.e.\ the classical paths. Quantum effects have their
origin in fluctuations around the classical paths. Therefore, it is useful to 
decompose a general path into the classical path and a fluctuation around it. 
Expanding the action in powers of the fluctuations the second order term yields 
the leading quantum corrections. Higher order terms are often neglected within 
a semiclassical approximation which becomes exact for linear systems.

In the previous section we have derived an effective equation of motion for the
system variable by eliminating the external degrees of freedom. The same
procedure may of course also be carried out within the path integral formalism
\cite{feynmanvernon63, grabertetal88,HTB90,weiss93}. The influence of the
environment is then contained in an effective action which has to be added to
the action of the system and which in imaginary time is given by
\cite{grabertetal88,HTB90,weiss93}
\begin{equation}
S_{eff}[q]=-\frac{1}{4}\int_0^{\hbar\beta}d\tau\int_0^{\hbar\beta}
d\sigma k(\tau-\sigma)[q(\tau)-q(\sigma)]^2
\label{eq:effact}
\end{equation}
where
\begin{equation}
k(\tau)=\frac{M}{\hbar\beta}\sum_{n=-\infty}^{+\infty}\vert\nu_n\vert
\hat\gamma(\vert\nu_n\vert)\exp(i\nu_n \tau)
\label{eq:kvtau}
\end{equation}
and $\hat\gamma(z)$ denotes the Laplace transform of the damping kernel
$\gamma(t)$. The effective action (\ref{eq:effact}) is clearly nonlocal
and can thus not be expressed in terms of a potential. If the potential
renormalization term in the Hamiltonian (\ref{eq:hamiltonian}) would be
absent, there would have been a local contribution in (\ref{eq:effact}).
The self\/interaction of the paths induced by (\ref{eq:effact}) via the
kernel (\ref{eq:kvtau}) decays for ohmic damping only algebraically as
$\tau^{-2}$ and therefore represents a long range interaction.

\subsection{Application: The dissipative decay of a metastable state}
\label{sec:decay}
\begin{figure}
\includegraphics[width=0.8\columnwidth]{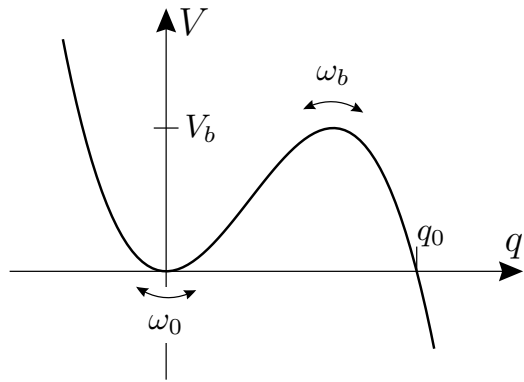}
\caption{Cubic potential as defined in Eq.~(\protect\ref{eq:cubicpot}).}
\label{fig:cubic}
\end{figure}

A local potential minimum may be metastable due to the environmental coupling
and quantum effects. Correspondingly, there are two escape mechanisms: thermal
activation which dominates at high temperatures and quantum tunnelling which
becomes important at low temperatures. To be definite, we consider the cubic
potential
\begin{equation}
V(q)=\frac{M}{2}\omega_0^2q^2\left(1-\frac{q}{q_0}\right)
\label{eq:cubicpot}
\end{equation}
which is depicted in Fig.~\ref{fig:cubic}. The barrier height is given by
$V_b=(2/27)M\omega_0^2q_0^2$ and, in this special case, the
barrier angular frequency $\omega_b$ equals the well angular frequency 
$\omega_0$. 

\begin{figure}
\includegraphics[width=0.8\columnwidth]{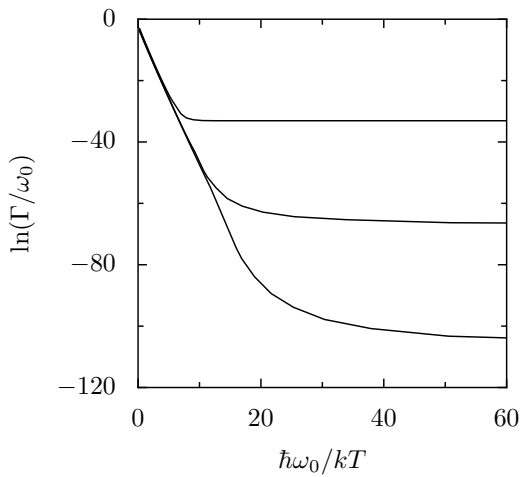}
\caption{Arrhenius plot for the decay rate of a metastable state. The
damping strength varies from the upper to the lower curve as
$\gamma/2\omega_0=0, 0.5,$ and 1 (data taken from Ref.~\onlinecite{GOW87}).}
\label{fig:rates}
\end{figure}

In Fig.~\ref{fig:rates}, the decay rate is shown in an Arrhenius plot.
At the so-called crossover temperature $T_0$, see Eq.~(\ref{eq:crosstemp})
below, there is a rather distinct
transition between the thermal regime on the left side and the quantum regime
on the right side \cite{ingoldPRL}. Furthermore, we observe that the thermal
regime is larger for stronger damping, i.e.\ the system becomes more classical.

While a real time approach to dissipative decay is feasible
\cite{ingolddiss,hofma91,anker95,anker99}, a simpler
alternative is provided by an imaginary time calculation where the partition 
function $Z_{\beta}$ is considered. Since the potential (\ref{eq:cubicpot}) is 
not bounded from below, it is no surprise that strictly speaking $Z_{\beta}$ 
does not exist. From the path integral point of view there exists an unstable 
fluctuation mode around the barrier which leads to a saddle point in function 
space. One can circumvent this difficulty by performing the integration in the 
direction of steepest descent. The partition function and as a consequence also 
the free energy then acquire an imaginary part which may be related to the 
decay rate \cite{caldeiraleggett83, HTB90}. For details of this relation we 
refer the reader to the discussion in Ref.~\onlinecite{hanggihontscha90}.

The transition between thermal and quantum regime can be well understood
within the path integral picture by considering the possible classical
paths of duration $\hbar\beta$ in the inverted cubic potential. For high
temperatures or short imaginary times $\hbar\beta$ the only classical solutions
are the constant solutions $q=0$ in the well and $q_{\rm b}=2q_0/3$
at the barrier. Below a temperature given by the positive solution of
\begin{equation}
\nu_1^2+\vert\nu_1\vert\hat\gamma(\vert\nu_1\vert)-\omega_b^2=0
\label{eq:ghhm}
\end{equation}
a second fluctuation mode becomes unstable, thereby indicating a new classical
solution which performs an oscillation around the barrier \cite{RHF85}. This 
new solution is associated with quantum tunnelling. Therefore, (\ref{eq:ghhm}) 
defines the crossover temperature which for ohmic damping is given by 
\cite{ingoldPRL}
\begin{equation}
T_0=\frac{\hbar}{2\pi k}\left[\left(\frac{\gamma^2}{4}
+\omega_b^2\right) ^{1/2}-\frac{\gamma}{2}\right].
\label{eq:crosstemp}
\end{equation}
As discussed above, stronger damping leads to a lower crossover temperature
and smaller quantum regime. It thus makes the system more classical. A distinct
feature of the dissipative quantum decay in the low temperature regime is its
algebraic enhancement of the decay rate with temperature \cite{HTB90}. For the
case of an ohmic environment with a constant friction behavior at low
frequencies one finds a universal $T^2$-enhancement of both, the prefactor and
the effective action, with the latter dominating the exponential rate
enhancement \cite{HTB90,grabertetal84}.

\section{Sundry remarks and conclusions}

With this work we elucidated the topic of quantum Brownian noise which drives
the dynamics of open dissipative quantum systems. We have emphasized the strong
implications that thermal equilibrium and time-reversal symmetry (leading to
detailed balance symmetry) imposes on the reduced system dynamics. We also
pointed out the advantageous use of the path integral scheme for the case of
nonlinearity and strong friction.

This method seemingly is superior to any perturbative scheme that treats the
system-bath coupling to low orders only, such as the weak coupling master
equation methodology \cite{grabert82,haake73,spohn80,alicki87}. There are
recent developments in the strong friction regime, where an alternative
description in terms of a quantum Smoluchowski equation is promising
\cite{ankerhold}, see also the contribution by Grabert, Ankerhold and Pechukas
in this special issue.

A consequent use of the so-called rotating-wave approximations also may
entail some danger. It safely can be applied only in the weak
coupling regime for resonant situations. We remark that the use of the
rotating-wave approximation
implies a violation of the Ehrenfest theorem in the order of $\gamma^2$,
\cite{talkner86,ford96} which is clearly small only in the weak coupling
regime, i.e.\ for $\gamma \ll \omega_0$ , with $\omega_0^{-1}$ denoting some
typical time scale of the system dynamics. The same remarks apply to the 
failure of the quantum regression theorem 
\cite{talkner86,grabert82b,fordoconnell96}: Again, the effect might be small 
for (i) very weak damping, (ii) not too low temperatures obeying 
$kT\gg \hbar\gamma$ and (iii) not too short evolution times.

The generalized quantum Langevin equation discussed in 
Sec.~\ref{sec:qmdissipative} is formally exact for nonlinear quantum systems. 
Its practical use is typically restricted, however, to linear systems for which 
the response can be evaluated in closed form. This holds true even for 
time-dependent linear systems for which the response is still linear although 
the evaluation involves the use of numerical Floquet theory 
\cite{zerbehanggi95}. The lack of knowledge of this generally nonlinear 
response function also plagues the evaluation of the corresponding generalized 
master equations.

This problem of obtaining the generalized master equation from the nonlinear 
generalized Langevin equation is not solved either for the classical problem 
with colored noise \cite{hanggi97}. It is also this very problem that limits 
the practical use of the various variants of recently derived stochastic 
Schr\"odinger approaches \cite{SS}.

Likewise, the use of nonlinear, but non-stochastic Schr\"odinger
equations of the type discussed and surveyed in
Refs.~\onlinecite{kostin75,messer79,dekker81} can clearly not
describe the time evolution of a quantum mechanical mixture, nor do these nonlinear
deterministic approaches obey, in general,
the Heisenberg uncertainty relation \cite{messer79,dekker81}.

There have been repeated attempts since the early days of quantum mechanics 
to explain quantum phenomena in terms of Einstein's theory of classical 
diffusion. Early efforts in this direction were those of
F\"urth \cite{furth33}, F\'enyes \cite{fenyes52}, Weizel \cite{weizel53}
and Favella \cite{favella67}. This credo has been popularized
later by Nelson \cite{nelson66, ghirardietal78} under the label of
``Stochastic Mechanics''.  It can  convincingly be demonstrated, however,
that a quantum dynamics is quite distinct from a
classical Markovian -- or even non-Markovian -- stochastic dynamics \cite{GHT79,wangliang93}.
This holds even more so, if one attempts to
incorporate the quantum dissipation for an open system.

These sundry remarks thus give clear evidence that the topic of quantum 
Brownian motion
-- although 100 years have passed since Einstein's cornerstone contribution
\cite{einstein1905}-- cannot be considered as ``solved''. For example, little
is presently  known also for the description and the role of quantum noise
acting in steady state, far from equilibrium situations, i.e.\ when several 
baths of different nature and/or different temperature
are coupled to the nonlinear system of interest.

The latter case is also of salient importance for the description
of the quantum dynamics of so-called quantum Brownian motors \cite{QBM}.
In those applications quantum Brownian
noise is utilized in combination with non-equilibrium (classical or quantum)
fluctuations to perform exploitable work against external bias forces. In 
summary, the field of quantum Brownian motion is very much alive and lots of 
challenges still need to be addressed and mastered.

\begin{acknowledgments}
PH gratefully acknowledges financial support by the DAAD-KBN
(German-Polish project \textit{Stochastic Complexity}), the Foundation
for Polish Science [Fundacja na Rzecz Nauki Polskiej], the Deutsche
Forschungsgemeinschaft via grant HA 1517/13-4 and the collaborative research
grants SFB~486 and SFB~631.
\end{acknowledgments}

\end{document}